\documentclass[12pt,footnotenumbering]{article}
\usepackage{amsmath}
\usepackage{amsbsy}
\usepackage{amsthm}
\usepackage{amsfonts}
\usepackage{amssymb}
\usepackage{graphicx}
\usepackage{epstopdf}
\usepackage{tikz}
\usepackage{color}
\usepackage{dsfont}
\usepackage{pgfplots}
\usepackage{subfig}
\usetikzlibrary{fit}
\usetikzlibrary{shapes.geometric}
\usetikzlibrary{decorations.pathmorphing}

\tikzset{vertex/.style={circle,fill=black,inner sep=2pt},
bigvertex/.style={circle,fill=black,inner sep=4pt},
specialEP/.style={rectangle,fill=white,draw,inner sep=3pt},
nuEP/.style={circle,fill=white,draw, inner sep=2pt},
linelabel/.style={sloped,above,very near start, inner sep=1pt,execute at begin node=$\scriptstyle,execute at end node=$},
baseline=(current  bounding  box.center),doubled/.style={double distance= 1pt,line width=1.5pt}
}
\pgfdeclarelayer{background}
\pgfsetlayers{background,main}

\numberwithin{equation}{section}

\let\a=\alpha \let\b=\beta    \let\g=\gamma     \let\d=\delta    
\let\e=\varepsilon
       \let\th=\theta     
\let\l=\lambda
\let\m=\mu

 \let\D=\Delta       \let\L=\Lambda    
\let\P=\Pi

\let\==\equiv

\def\media#1{{\left\langle#1\right\rangle}}

\def\xx{{\bf x}}
\def\yy{{\bf y}}\def\zz{{\bf z}}

\def\be{\begin{equation}}
\def\ee{\end{equation}}
\def\bea{\begin{eqnarray}}\def\eea{\end{eqnarray}}

\def\1{{\mathds 1}}

\pgfdeclarelayer{background}
\pgfsetlayers{background,main}

\begin{document}

\title{Columnar Phase in Quantum Dimer Models}

\author{\vspace{5pt} Alessandro Giuliani$^{1}$ and Elliott H.~Lieb$^{2}$\\
  \vspace{-4pt}\small{$^{1}$Dipartimento di Matematica e Fisica,  Universit\`a di Roma
Tre} \\ \small{
L.go S. L. Murialdo 1, 00146 Roma, Italy}\\
  \vspace{-4pt}\small{$^{2}$Departments of Mathematics and Physics,
    Jadwin Hall, Princeton University}\\
\small{    Princeton 08544 NJ, USA}}

\date{April 29, 2015}

\maketitle

\begin{abstract}
The quantum dimer model, relevant for short-range resonant valence bond
physics, is rigorously shown to have long range order in a cr{y}stalline
phase in the attractive case at low temperature
and not too large flipping term. This term flips horizontal dimer pairs
to vertical pairs (and vice versa) and is responsible for the word
`quantum' in the title. In addition to the
dimers, monomers are also allowed.
The mathematical method used is `reflection positivity'. The model {and proof} can
easily be generalized to dimers or  plaquettes in
3-dimensions.

\end{abstract}

\renewcommand{\thefootnote}{$ $}
\footnotetext{\copyright\, 2015 by the authors. This paper may be reproduced, in its
entirety, for non-commercial purposes.}

\renewcommand{\thesection}{\arabic{section}}

\section{Introduction}

Starting with  the  work of Rokhsar and Kivelson \cite{RK} quantum dimer
models became popular in connection with the study of resonating valence
bond (RVB) states, as well as in the study of cold bosons, frustrated
magnetism, Josephson junction arrays, and other physical models, see
\cite{MR} for a review. In these
models dimers on a square lattice not only have a pair interaction
that favors or disfavors parallel dimers next to each other, but {they} also 
{have} a flipping term that exchanges a horizontal pair with a vertical pair. 
It is this `non-diagonal' flipping term that gives rise to the appellation
`quantum'. 
 
Many kinds of physical phases can arise from such models. {As the interaction is increased from large negative values (strongly  attractive case) to large positive values (strongly repulsive case),
quantum dimers are expected 
to exhibit a remarkable series of transitions: for instance, on the square lattice, 
the system is expected to pass from a crystalline ``columnar" state, to a ``mixed" state, then a ``plaquette", and finally a ``staggered" state, passing through 
an anomalous liquid phase at the Rokshar-Kivelson point, see \cite{MR,RPM} for details about these phases and transitions. Up to now, none of these conjectured phases and transitions has been 
rigorously established. For recent rigorous results about the anomalous liquid phase in the {\it classical} 
interacting dimer model, see \cite{GMT1,GMT2}. 
}

Here we study
the attractive case on the square lattice and rigorously establish that
there is a columnar
state at low temperature provided the monomer density and flipping rate
are not too large. Our analysis easily extends to the  three-dimensional
cubic lattice, in which case the dimers can remain as dimers or they
can be replaced by two-dimensional plaquettes. 

Much before Ref.\cite{RK}, Heilmann and Praestgaard \cite{H, HP, HP2}, using
a Peierls contour argument, 
proved that
the same model, but without the flipping term, has columnar long-range
order (LRO) at low temperature and not too large monomer density. 

The introduction of the quantum feature complicates the proof of LRO
considerably. Borgs et al and Datta et al \cite{BKU, DFF} proved, quite
generally, that quantum perturbations of classical models do not destroy
LRO at low temperature. Their methods are both based on a quantum extension
of Pirogov-Sinai theory \cite{PS}.

In our case the method we use -- reflection positivity (RP) and chessboard
estimates -- is much simpler. While it was introduced in quantum field
theory by Osterwalder and Schrader \cite{OS}, its first application in
classical statistical mechanics goes back to Fr\"ohlich, Simon and Spencer
\cite{FSS}, and to Dyson, Lieb and Simon for the quantum case \cite{DLS}.
See also \cite{FL} for an application to the 
anisotropic quantum  Heisenberg models, which is relevant for our proof
here. The use of RP has
added advantages apart from brevity. When applicable it probably gives
better estimates on the range in which LRO holds. {In \eqref{main} we give an explicit sufficient condition for LRO.} It also establishes some
correlation positivity results, which can be used for
some correlation inequalities.

\section{The model}

We introduce a quantum dimer model on the 2D
square lattice {(which can be easily extended to 3D)}: consider a square portion of 
$\mathbb Z^2$ of even side $L$ with periodic boundary conditions (for
convenience later we will choose $L$ to be divisible by 4). The
Hamiltonian is conventionally expressed in the following way: 
\be\label{ham}
H_\L=-\e\sum_{P\subset\L}(v_P^\dagger h_P+h_P^\dagger v_P)-\mu
\sum_{P\subset\L} (v_P^\dagger v_P+h_P^\dagger h_P)+zM\equiv -\e T-\mu U+zM,
\ee
where the summations run over the plaquettes $P$ of $\L$. 
The physical objects on the lattice are dimers, which lie on the edges, 
and the symbols $h_P$ (resp. $v_P$) denote operators that kill two
horizontal (resp. vertical) dimers occupying $P$. Correspondingly,
$h_P^\dagger$ and $v_P^\dagger$ create pairs of parallel dimers. Of 
course $P$ can have no dimers, or one dimer, in which
case $h_P$ and $v_P$ annihilate the corresponding state; {this means that $h_P|\phi\rangle=
v_P|\phi\rangle=0$ if $|\phi\rangle$ is a classical dimer configuration such that $P$ has zero or one dimers.} We
{\it do not} assume complete dimer covering: if $N_v$ and $N_h$
denote the number of dimers in the vertical and horizontal directions,
$M=|\L|-2N_v-2N_h$ denotes the number of monomers (empty sites).

The first term in \eqref{ham} switches the dimers in a doubly occupied
plaquette, while the second simply counts the number of such plaquettes. 3D extensions of the model,
treatable by our methods, are, e.g., the following: (i) same Hamiltonian in 3D, where now plaquettes can be in three possible directions; (ii) same thing, but with the interaction term replaced by a similar one, counting the number of unit cubes completely filled by four parallel dimers; (iii) quantum Hamiltonian for plaquettes (rather than dimers): 
parallel neighboring plaquettes interact and can flip whenever they occupy the same unit cube. See section \ref{sec3}.

The sign of the constant $\e$ is conventional, because it can be changed by
the unitary transformation $i^{N_v}$.
Therefore, we will assume that $\e>0$. The sign of the second term 
is very important: we shall be concerned with the case that $\m>0$. By
rescaling, we assume $\m=1$. In these units, we assume that the
monomer chemical potential, $z$, is larger than $-1/2$. 

The Hamiltonian preserves the number of dimers (or monomers) and this
divides the Hilbert space into sectors, each of which has a spectrum.
At $\e=0$, there are four ground states, called columnar states, and
these are in the sector where the number of dimers is $|\L|/2$. One of them
consists of parallel columns of horizontal dimers, and the other three are
obtained by shifting the columns one lattice space to the right, or
else rotating the whole picture(s) by $90^o$.

As anticipated above, we prove that columnar LRO, in the sense defined
precisely in \eqref{lro}, is present at small
enough $\e>0$ and large enough $\b$, by using quantum reflection positivity,
chessboard
estimates, 
exponential localization and an adaptation of the classical Peierls
argument. 

More exactly, we prove that a sufficient condition for LRO is that
the following inequality is satisfied by the three parameters $\b,
\e$ and $z$:
\be \boxed{ \sum_{n\ge 2}n^2\a^n<{\frac{3}{7}}\,\quad{\rm with}\quad 
\a=18\max\Big\{e^{-\b(\frac1{16}-\frac{\e}2)}+\big(\frac{2\e}{1/16-\e}
\big)^ {\tfrac1{24}},\,
e^{-\frac12\b(z+1/2)}\Big\}}\label{main}
\ee

\subsection{Reflection positivity}

We first show that the Hamiltonian \eqref{ham} in a suitable representation
is reflection positive. We associate with each point $\xx\in\L$ a Hilbert
space $\mathcal H_\xx=\mathbb C^5$; there are 5 allowed states at $\xx$,
one of them is the monomer, and the other 4 refer to the
half-dimers pointing in the directions north, east, south, west. The total
Hilbert space is the tensor product of these local spaces. To ensure that 
we do not have unpaired half-dimers, we will add to our Hamiltonian a term 
that suppresses unpaired half-dimers. At each point we introduce the
operators $\mathds 1^\#_\xx$, with $\#\in\{0,N,E,S,W\}$, which are
projection operators onto the 5 different states, the rotation operator
$R^{N\to E}_\xx$, which flips the state $N$ into $E$ and gives zero
otherwise, and similarly $R^{N\to W}_\xx$, etc. 

Consider the Hamiltonian:
\bea H^\l_\L&=&\l\Big[|\L|-2\sum_{\xx}(\mathds 1^E_\xx \mathds
1^W_{\xx+\hat e_1}+
\mathds 1^N_\xx \mathds 1^S_{\xx+\hat e_2})-\sum_\xx\mathds
1^0_\xx\Big]+z\sum_\xx\1^0_\xx\label{2.2} \\
&&-\e \sum_{P=(\xx_1,\xx_2,\xx_3,\xx_4)}(R^{E\to
N}_{\xx_1}R^{W\to N}_{\xx_2}R^{W\to S}_{\xx_3}R^{E\to
S}_{\xx_4}+h.c.)\label{2.3}\\
&&-\sum_{P=(\xx_1,\xx_2,\xx_3,\xx_4)}(\1_{\xx_1}^N\1_{\xx_2}^N\1_{\xx_3}
^S\1^{S}_{\xx_4}+\1_{\xx_1}^E\1_{\xx_2}^W\1_{\xx_3}
^W\1^{E}_{\xx_4}) \label{2.4}
\eea
where the sums in the last two lines runs over the plaquettes of $\L$, 
each plaquette $P$ being thought as the union of the four sites
$\xx_1,\xx_2,\xx_3,\xx_4$ at the left bottom, right bottom, right top, left
top of $P$, respectively. As we let $\l\to\infty$, $H^\infty_\L$ gives
infinite energy to unphysical configurations (i.e., configurations with
unpaired half-dimers) and on the physical space it becomes equivalent to
\eqref{ham} at $\m=1$: note, in fact, that the term in square brackets on
the first line is zero on physical configurations and positive otherwise; 
the remaining terms \eqref{2.3} and \eqref{2.4} reduce on the physical space
to the first and second terms in \eqref{ham}, and similarly
$z\sum_\xx\1^0_\xx$ reduces to the last term in \eqref{ham}.

In this representation, there is a natural notion of reflection positivity.
Take a pair of planes passing through the bonds and cutting $\L$ in two
equal sized halves, $\L_-$ and $\L_+$. {Reflection} about these planes
{consists} in changing the site labels into their reflected images, and
exchanging
up with down ($N$ with $S$) or right with left ($E$ with $W$), depending on
whether the planes are horizontal or vertical. We call $\th$ such a
reflection; it satisfies the conditions listed in \cite{FILS}. Given any
pair of reflection planes, we can write $-H^\l_\L$ as 
$A+\th A+\sum_i C_i\th C_i$, with $A$ representing the restriction of
$-H^\l_\L$ to one of the two halves, say $\L_+$, and $C_i\th C_i$
being the terms connecting the two halves. Therefore, $H^\l_\L$ is
Reflection Positive (RP), which follows from the general theory in
\cite[Sections 2 and 3.3]{FILS}. This means that if we take any operator $F$
supported on $\L_+$ then
\be {\media{F^\dagger(\th F)}_\l}\ge 0,\label{rp}\ee
where $\media{(\cdot)}_{{\l}}={\rm Tr}e^{-\b
H^\l_\L}(\cdot)/Z_\L^\l$,
with $Z_\L^\l={\rm Tr}e^{-\b H^\l_\L}$. Moreover, 
the chessboard estimate holds, see
\cite[Theorem 4.1]{FILS}. If, e.g., $\{f_\xx\}_{\xx\in\L}$ are
operators supported on the sites of $\L$
\be \big\langle{\prod_{\xx\in\L}f_\xx}\big\rangle_{{\l}}\le
\prod_{\xx\in\L}\big[\media{F_\xx}_{{\l}}\big]^{1/|\L|},\label{chess}\ee
where
$F_\xx=\prod_{\zz\in\L}\tilde f_\xx^{(\zz)}$, and 
$\tilde f_\xx^{(\zz)}$ is a copy of $f_\xx$ 
attached to the site $\zz$ if $\zz$ and $\xx$ belong to the same  
colored sublattice on the chessboard. If they belong to differently colored
sublattices, then $\tilde f_\xx^{(\zz)}$ is a copy of $\th f_\xx$ 
attached to the site $\zz$. Eq.\eqref{chess} is obtained by repeatedly 
reflecting the left side about planes not passing through the sites.

Similar considerations apply to the case in
which we have to evaluate the average of a product of operators supported
on the occupied bonds of one of the four classical (columnar) ground
states. In this case we repeatedly
reflect about planes that pass neither through sites, nor through
the occupied bonds. For these kind of reflections, we need $L$ to be
divisible by 4. In the next {section}  we will apply this analogue of
\eqref{chess} to estimate the probability of any
Peierls contour in terms of that of the `universal Peierls contour' that is
as large as $\L$ itself, and is much easier to evaluate. 

Note that \eqref{rp} and \eqref{chess} remain valid as $\l\to\infty$, in
which case the average $\media{\cdot}_{{\l}}$ reduces to the Gibbs measure of
the original quantum dimer model, simply denoted by $\media{\cdot}$. From
now on we restrict ourselves to this limiting case, and we implicitly assume
that no unphysical configuration of half-dimers appears in the states
under consideration. We also drop the $\infty$ label from $H_\L^\infty$ and
identify $H_\L$ with $-U-\e T+zM$, as in \eqref{ham} (recall that we fixed
$\m=1$).

\subsection{The Peierls argument}

We introduce on-site projectors associated with the 4 classical
($\e=0$) ground states, labelled $1,2,3,4$, as follows:
\bea && \1^1_\xx=\begin{cases} \1^E_\xx\ {\rm if}\ x_1\ {\rm is}\ {\rm
even}\\
\1^W_\xx\ {\rm if}\ x_1\ {\rm is}\ {\rm
odd}\end{cases},\qquad 
\1^2_\xx=\begin{cases} \1^W_\xx\ {\rm if}\ x_1\ {\rm is}\ {\rm
even}\\
\1^E_\xx\ {\rm if}\ x_1\ {\rm is}\ {\rm
odd}\end{cases},\\
&& \1^3_\xx=\begin{cases} \1^N_\xx\ {\rm if}\ x_2\ {\rm is}\ {\rm
even}\\
\1^S_\xx\ {\rm if}\ x_2\ {\rm is}\ {\rm
odd}\end{cases},\qquad 
\1^4_\xx=\begin{cases} \1^S_\xx\ {\rm if}\ x_2\ {\rm is}\ {\rm
even}\\
\1^N_\xx\ {\rm if}\ x_2\ {\rm is}\ {\rm
odd}\end{cases}.
\eea
We want to prove Long Range Order (LRO), in the sense that 
\be \sum_{i=1}^4\media{\1_\xx^i \1_{\yy}^i}>{1/4},\label{lro}\ee
for all $\L,\xx,\yy$. {The condition \eqref{lro} guarantees that the infinite volume Gibbs state 
$\media{\cdot}_{\mathbb Z^2}=\lim_{|\L|\to\infty}\media{\cdot}$
is not pure, i.e., it does not satisfy the clustering property$^1$\footnote{$^1$ The clustering property is the condition that $\lim_{|\xx-\yy|\to\infty}
\media{A_\xx B_\yy}_{\mathbb Z^2}=\media{A_\xx}_{\mathbb Z^2}\media{B_\yy}_{\mathbb Z^2}$ 
for any two observables $A_\xx$ and $B_\yy$ supported around $\xx$ and $\yy$, respectively.}: therefore, \eqref{lro} implies that  the model exhibits a phase transition, in that it admits multiple pure states (at least four, by the translational and rotational symmetries). To see that \eqref{lro} violates the clustering property, note that $\media{\1_\xx^i}=(1-p_0)/4$, $\forall \xx\in\L, \forall i=1,2,3,4$, where $p_0=\media{\1^0_\xx}$. The clustering property would imply $\lim_{|\xx-\yy|\to\infty}\sum_{i=1}^4\media{\1_\xx^i \1_{\yy}^i}=
(1-p_0)^2/4\le1/4$, which is in contradiction with \eqref{lro}.
}

Recalling that 
$\sum_{i,j=0}^4\media{\1^i_\xx\1^j_\yy}=1$, 
a sufficient
condition for \eqref{lro} is that for small enough temperature and monomer
density, 
\bea &\media{\1_\xx^i \1_{\yy}^j}<{\frac1{21}\cdot\frac34=\frac1{28}},&{\rm if}\ 0<i,j\le 4\ {\rm and}\
i\neq j,\label{2.7}\\
&
\media{\1^0_\xx \1^i_\yy}\le
\media{\1^0_\xx}<{\frac1{28}},& \forall i=0,\ldots,4,\label{2.9}\eea 
for all $\L,\xx,\yy$. The number $21=25-4$ is the number of possible
`bad pairs' of indices attached to the projectors in $\xx$ and $\yy$.

In order to bound the left side of \eqref{2.7}, we insert at each site $\zz$
different from $\xx$ and $\yy$ a resolution of the identity $I=\sum_{i=0}^4
\1^i_\zz$, and then we expand the product of sums, thus getting a sum over
{\it configurations} of projectors (we repeat that, since we let
$\l\to\infty$, only physical configurations are retained). These
configurations are the quantum analogues of the classical dimer
configurations in \cite{HP}. In analogy with the classical case, we
associate each configuration with a set of Peierls contours \cite{H,HP,HP2},
which are closed paths on 
the dual lattice, separating either good regions of different types (a
good region being a connected region covered by projectors all with the same
index $i\in\{1,2,3,4\}$) or a good region from a bad region. We
identify the contour enclosing the good region containing $\xx$ and call it
$\g$. Every bond {\it crossing} $\g$ comes with two projectors associated
with a bad pair of indices. These bonds can be of 4 different types,
depending on whether they are of the form $(\xx,\xx+\hat e_1)$ with $x_1$
even, $(\xx,\xx+\hat e_1)$ with $x_1$
odd, $(\xx,\xx+\hat e_2)$ with $x_2$
even, or $(\xx,\xx+\hat e_2)$ with $x_2$
odd. We single out the most numerous among these four types, which consist,
therefore, of at least $|\g|/4$ elements; let $T_\g$ be the corresponding
set of bonds. 

For the purpose of an upper bound, we retain only the projectors on the
bonds in $T_\g$ and throw away (i.e.,
bound by 1) all the others. As a result,
\be \media{\1_\xx^i \1_{\yy}^j}\le \sum_{\g\,\succ\, \xx}\ 
\sum_{\{(i_\zz,i_{\zz'})\}_{(\zz,\zz')\in
T_\g}}\Big\langle{\prod_{(\zz,\zz')\in T_\g}P^{i_\zz}_\zz
P^{i_{\zz'}}_{\zz'}}\Big\rangle,\label{peierls}\ee
where the first sum runs over contours enclosing $\xx$ and the second over
the bad pairs of indices associated with the bonds $(\zz,\zz')$ in
$T_\g$; if $\zz$ is the site belonging to the same good region as $\xx$,
then $i_\zz=i$.

In order to evaluate $\langle{\prod_{(\zz,\zz')\in
T_\g}P^{i_\zz}_\zz P^{i_{\zz'}}_{\zz'}}\rangle$, 
we use the chessboard estimate 
\cite[Theorem 4.1]{FILS}, in the way explained at the end of previous
section, thus finding 
\be \Big\langle{\prod_{(\zz,\zz')\in
T_\g}P^{i_\zz}_\zz P^{i_{\zz'}}_{\zz'}}\Big\rangle\le
\d^{|T_\g|},\label{bound} \ee
where 
\be
\d^{|\L|/2}=\max\Big\{\media{P^{34}},\media{P^{32}},\media{P^{30}},
\media { P^ { 20 } } \Big\}\ee
and 
\be P^{ij}=\prod_{\xx:\ x_1\in 4\mathbb Z}
\1^i_\xx\1^j_{\xx+\hat e_1}\1^j_{\xx+2\hat e_1}\1^i_{\xx+3\hat
e_1}\label{up}\ee
is a {\it universal projector}, in the sense of \cite{FL}, onto a
particular periodic classical state. The cases of interest, that is
$P^{34}, P^{32}, P^{30}$ and $P^{20}$, are described in Figure 1. Each
one involves all the vertices in $\L$, not just those in the bonds of
$T_\g$.

\begin{figure}[h]
\centering
\begin{tikzpicture}[line width=2pt]
\node[] at (1.8,-1.2) {$P^{34}$};
\node[] at (8.1,-1.2) {$P^{32}$};
\node[] at (1.8,-6.8) {$P^{30}$};
\node[] at (8.1,-6.8) {$P^{20}$};

\begin{scope}[scale=0.7]
\draw[lightgray,line width=0] (-1,-1) grid ++(7,5);

\draw[line width=3] (-1,3) -- (-1,4);
\draw[line width=3] (-1,-1) -- (-1,0);
\draw[line width=3] (-1,1) -- (-1,2);
\draw[line width=3] (0,3) -- (0,4);
\draw[line width=3] (0,-1) -- (0,0);
\draw[line width=3] (0,1) -- (0,2);

\begin{scope}[shift={(0,2)}]
\draw[line width=3] (1,3) -- (1,2);
\draw[line width=3] (1,-1) -- (1,-2);
\draw[line width=3] (1,1) -- (1,0);
\draw[line width=3] (2,3) -- (2,2);
\draw[line width=3] (2,-1) -- (2,-2);
\draw[line width=3] (2,1) -- (2,0);
\end{scope}

\draw[line width=3] (3,3) -- (3,4);
\draw[line width=3] (3,-1) -- (3,0);
\draw[line width=3] (3,1) -- (3,2);
\draw[line width=3] (4,3) -- (4,4);
\draw[line width=3] (4,-1) -- (4,0);
\draw[line width=3] (4,1) -- (4,2);

\begin{scope}[shift={(0,2)}]
\draw[line width=3] (5,3) -- (5,2);
\draw[line width=3] (5,-1) -- (5,-2);
\draw[line width=3] (5,1) -- (5,0);
\draw[line width=3] (6,3) -- (6,2);
\draw[line width=3] (6,-1) -- (6,-2);
\draw[line width=3] (6,1) -- (6,0);
\end{scope}

\begin{scope}[shift={(1,0)}]
\draw[lightgray,line width=0] (7,-1) grid ++(7,5);
\draw[line width=3] (7,3) -- (7,4);
\draw[line width=3] (7,-1) -- (7,0);
\draw[line width=3] (7,1) -- (7,2);
\draw[line width=3] (8,3) -- (8,4);
\draw[line width=3] (8,-1) -- (8,0);
\draw[line width=3] (8,1) -- (8,2);

\draw[line width=3] (9,3) -- (10,3);
\draw[line width=3] (9,4) -- (10,4);
\draw[line width=3] (9,2) -- (10,2);
\draw[line width=3] (9,1) -- (10,1);
\draw[line width=3] (9,0) -- (10,0);
\draw[line width=3] (9,-1) -- (10,-1);

\begin{scope}[shift={(8,0)}]
\draw[line width=3] (3,3) -- (3,4);
\draw[line width=3] (3,-1) -- (3,0);
\draw[line width=3] (3,1) -- (3,2);
\draw[line width=3] (4,3) -- (4,4);
\draw[line width=3] (4,-1) -- (4,0);
\draw[line width=3] (4,1) -- (4,2);
\end{scope}

\begin{scope}[shift={(4,0)}]
\draw[line width=3] (9,3) -- (10,3);
\draw[line width=3] (9,4) -- (10,4);
\draw[line width=3] (9,2) -- (10,2);
\draw[line width=3] (9,1) -- (10,1);
\draw[line width=3] (9,0) -- (10,0);
\draw[line width=3] (9,-1) -- (10,-1);
\end{scope}
\end{scope}

\begin{scope}[shift={(0,-8)}]
\draw[lightgray,line width=0] (-1,-1) grid ++(7,5);
\draw[line width=3] (-1,3) -- (-1,4);
\draw[line width=3] (-1,-1) -- (-1,0);
\draw[line width=3] (-1,1) -- (-1,2);
\draw[line width=3] (0,3) -- (0,4);
\draw[line width=3] (0,-1) -- (0,0);
\draw[line width=3] (0,1) -- (0,2);

\draw (1,-1) node[vertex]{};
\draw (1,0) node[vertex]{};
\draw (1,1) node[vertex]{};
\draw (1,2) node[vertex]{};
\draw (1,3) node[vertex]{};
\draw (1,4) node[vertex]{};
\begin{scope}[shift={(1,0)}]
\draw (1,-1) node[vertex]{};
\draw (1,0) node[vertex]{};
\draw (1,1) node[vertex]{};
\draw (1,2) node[vertex]{};
\draw (1,3) node[vertex]{};
\draw (1,4) node[vertex]{};
\end{scope}

\begin{scope}[shift={(4,0)}]
\draw (1,-1) node[vertex]{};
\draw (1,0) node[vertex]{};
\draw (1,1) node[vertex]{};
\draw (1,2) node[vertex]{};
\draw (1,3) node[vertex]{};
\draw (1,4) node[vertex]{};
\begin{scope}[shift={(1,0)}]
\draw (1,-1) node[vertex]{};
\draw (1,0) node[vertex]{};
\draw (1,1) node[vertex]{};
\draw (1,2) node[vertex]{};
\draw (1,3) node[vertex]{};
\draw (1,4) node[vertex]{};
\end{scope}
\end{scope}

\draw[line width=3] (3,3) -- (3,4);
\draw[line width=3] (3,-1) -- (3,0);
\draw[line width=3] (3,1) -- (3,2);
\draw[line width=3] (4,3) -- (4,4);
\draw[line width=3] (4,-1) -- (4,0);
\draw[line width=3] (4,1) -- (4,2);

\begin{scope}[shift={(1,0)}]
\draw[lightgray,line width=0] (7,-1) grid ++(7,5);

\begin{scope}[shift={(8,0)}]
\draw (1,-1) node[vertex]{};
\draw (1,0) node[vertex]{};
\draw (1,1) node[vertex]{};
\draw (1,2) node[vertex]{};
\draw (1,3) node[vertex]{};
\draw (1,4) node[vertex]{};
\begin{scope}[shift={(1,0)}]
\draw (1,-1) node[vertex]{};
\draw (1,0) node[vertex]{};
\draw (1,1) node[vertex]{};
\draw (1,2) node[vertex]{};
\draw (1,3) node[vertex]{};
\draw (1,4) node[vertex]{};
\end{scope}

\begin{scope}[shift={(4,0)}]
\draw (1,-1) node[vertex]{};
\draw (1,0) node[vertex]{};
\draw (1,1) node[vertex]{};
\draw (1,2) node[vertex]{};
\draw (1,3) node[vertex]{};
\draw (1,4) node[vertex]{};
\begin{scope}[shift={(1,0)}]
\draw (1,-1) node[vertex]{};
\draw (1,0) node[vertex]{};
\draw (1,1) node[vertex]{};
\draw (1,2) node[vertex]{};
\draw (1,3) node[vertex]{};
\draw (1,4) node[vertex]{};
\end{scope}
\end{scope}
\end{scope}

\begin{scope}[shift={(-2,0)}]
\draw[line width=3] (9,3) -- (10,3);
\draw[line width=3] (9,4) -- (10,4);
\draw[line width=3] (9,2) -- (10,2);
\draw[line width=3] (9,1) -- (10,1);
\draw[line width=3] (9,0) -- (10,0);
\draw[line width=3] (9,-1) -- (10,-1);

\begin{scope}[shift={(4,0)}]
\draw[line width=3] (9,3) -- (10,3);
\draw[line width=3] (9,4) -- (10,4);
\draw[line width=3] (9,2) -- (10,2);
\draw[line width=3] (9,1) -- (10,1);
\draw[line width=3] (9,0) -- (10,0);
\draw[line width=3] (9,-1) -- (10,-1);
\end{scope}
\end{scope}
\end{scope}
\end{scope}

\end{scope}
\end{tikzpicture}
\caption{A pictorial representation of the four universal projectors that
are obtained after repeated reflections of the bad bonds crossing the
contours. See
\eqref{bound} {\it et seq}. In the third and fourth pictures the dots
represent monomers.}
\label{fig1}
\end{figure}
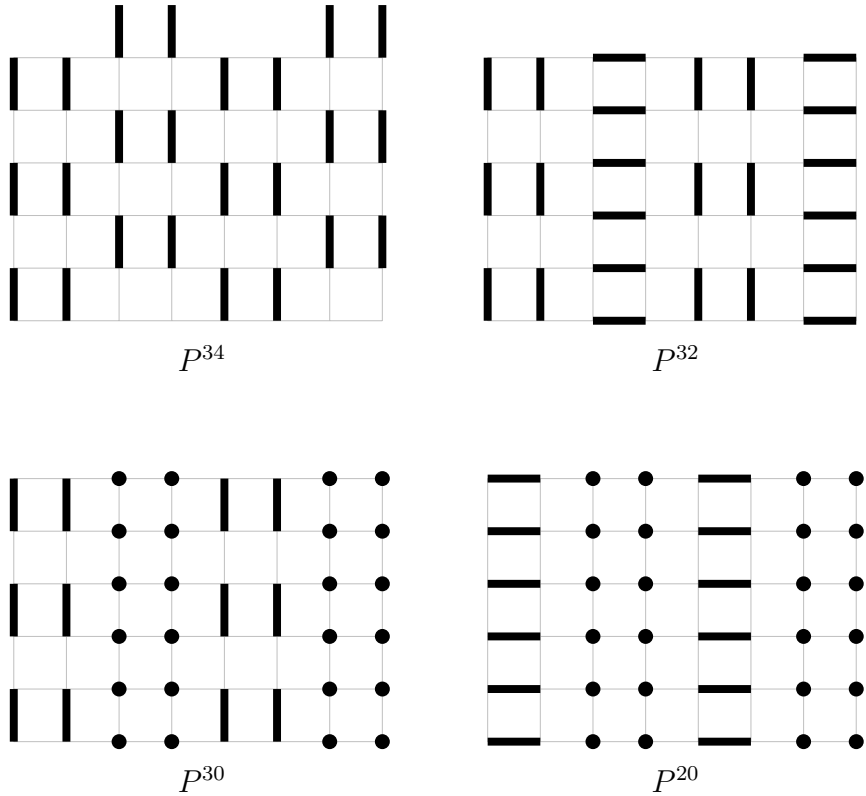

Similarly, the left side of \eqref{2.9} can be bounded by
$\media{P^0}^{1/|\L|}$, with
$P^0=\prod_{\xx\in\L} \1^0_\xx$. For later use, we also define
$P^i=\prod_{\xx\in\L} \1^i_\xx$. 

We will prove that \be\media{P^0}^{1/|\L|}\le e^{-\b(z+1/2)},\label{p0}\ee
which implies 
\eqref{2.9} for $z>-1/2$ and $\b$ large enough. Moreover, 
\be
\d\le\max\Big\{e^{-\b(1/8-\e)}+\big(\frac{2\e}{1/16-\e}\big)^{1/12},
e^{-\b(z+1/2)}\Big\}\label{d}
\ee
which is as small as desired, provided $z>-1/2$, and $\e$ and $\b^{-1}$
are sufficiently small. 

Assuming this, the validity of \eqref{2.7} and, therefore, existence of LRO,
follows
from the standard Peierls estimate: by \eqref{peierls}, \eqref{bound}, we get that if $i,j>0$ and $i\neq j$, and
$\d$ is small enough,
\be \media{\1_\xx^i \1_{\yy}^j}\le \sum_{\g\,\succ\, \xx}(4\d)^{|\g|/4} \ee
Now, the number of contours of length $2n$ (note that the length
of each contour is even) enclosing  $\xx$ is smaller than
$n^29^n/12$, which gives $\media{\1_\xx^i \1_{\yy}^j}\le{\frac1{12}}\sum_{n\ge
2}n^29^n(4\d)^{n/2}$. Requiring this to be {$<1/28$} gives \eqref{main}, provided that \eqref{p0}
and \eqref{d} are valid. Our
next goal is to prove these estimates!

\subsection{The universal projector}

We first illustrate the estimate for $\media{P^{32}}$, the ones for
$\media{P^{34}}$, $\ldots$, $\media{P^0}$ being very similar, or simpler. We
proceed
as in \cite[Section I.E and III]{FL}. Let $\{\psi_i\}_{i\ge 0}$ be an
orthonormal set of eigenfunctions of $H_\L$, numbered in order of
increasing energy (i.e., $E_0$ is the ground state energy). Moreover, let 
$\phi$ be the classical state corresponding to the universal projection
$P^{32}$, i.e., $P^{32}=|\phi\rangle\langle\phi|$. We
write 
\bea  \media{P^{32}}&=&\frac1{Z_\L} \sum_{i\ge 0}e^{-\b
E_i}|\media{\psi_i|\phi}|^2\label{2.15}\\
&=&\frac1{Z_\L}\Big[
\sum_{i:\ E_i<E_0+\D|\L|}e^{-\b
E_i}|\media{\psi_i|\phi}|^2+
\sum_{i:\ E_i\ge E_0+\D|\L|}e^{-\b
E_i}|\media{\psi_i|\phi}|^2\Big]\nonumber \\
&\equiv& R_-+R_+,\nonumber\eea
where $\D$ is a cutoff, which can be fixed to be $\D=1/16$: the criterion 
for the choice of $\D$ is that the energy $E_0+\D|\L|$ should be
approximately in the middle between the ground state energy $E_0$ and 
the energy of $\phi$. Neglecting $\e$, the ground state energy
would be $-|\L|/2$, while the energy of $\phi$ would be $-3|\L|/8$, i.e., it
would be separated from the ground state energy by a gap $2\D|\L|$ with
$\D=1/16$.

\subsubsection{Estimate of $R_+$}

We start with the easier term, $R_+$. Using the condition 
$E_i\ge E_0+\D|\L|$ we get 
\be R_+\le \frac1{Z_\L}e^{-\b(E_0+\D|\L|)}.\ee
To get a bound on $Z_\L$, we restrict the trace to one of the (monomer-free)
classical ground states, $D_0$. 
Thus,
\be Z_\L\ge \media{D_0|e^{\b(U+\e T-z M)}|D_0}\ge e^{
\b\media{D_0|(U+\e T-zM)|D_0}}=e^{\b U_0}\label{zeta}\ee
with $U_0=|\L|/2$. Moreover, $T\le U$, so that $H_\L\ge -U(1+\e)+z
M$
and, therefore, $E_0\ge -U_0(1+\e)$. All in all we find
\be R_+\le e^{-\b(\D|\L|-\e U_0)}=e^{-\b|\L|(1/16-\e/2)}\;.\ee

\subsubsection{Estimate of $R_-$}

We now turn to the more subtle $R_-$ term. Here we use the exponential
localization strategy of \cite{FL}. We bound
\be R_-\le \max_{i: E_i< E_0+\D|\L|}|\media{\psi_i|\phi}|^2\;.\ee
The goal is to show that the overlap between the two states is
exponentially small. It is clear that if $\e$ were zero then
$\phi$ and $\psi_i$ would be orthogonal because they would belong to
subspaces with very different energies. If $\e\neq 0$ is small, as is
the case here, the two states are eigenfunctions of different but close
Hamiltonians (with and without $\e T$) with very different energies. The
localization principle says that they are still almost orthogonal. 

Let $A=-U+U_0+zM$ and $B=-\e T$. We also need the projector $\P$ onto the 
subspace of eigenvectors of $A$ with eigenvalues larger than
$\frac32\D|\L|$. Note in particular that $\phi$ is in the range of $\P$. 
The eigenvalue equation for $\psi_i$ reads
\be (A+B)\psi_i=(E_i+U_0)\psi_i\equiv \l\psi_i,\ee
so that $\psi_i=-(A-\l+i\d)^{-1}(B-i\d)\psi_i$, for some $\d\ge 0$.
Therefore, $\media{\psi_i|\phi}=-\media{\psi_i|(B-i\d)(A-\l+i\d)^{-1}\phi}$
and, passing to the limit $\d\to 0$,
\be
\media{\psi_i|\phi}=-\langle\psi_i|B({A-\l})^{-1}|\phi\rangle.\label{2.20}
\ee
Not only $\phi$ is in the range of $\P$, but this is also the case for
$B({A-\l})^{-1}\phi$. Therefore, in \eqref{2.20} we can freely add some
projectors:
\be
\media{\psi_i|\phi}=-\langle\psi_i|\P B\P({A-\l})^{-1}|\phi\rangle.
\label{2.20d}
\ee
We can now iterate this, as long as $(B({A-\l})^{-1})^n\phi$ is in the
range of $\P$, which is the case if, e.g., $n\le \D|\L|/3$. We thus get 
\be
\media{\psi_i|\phi}=(-1)^n\langle\psi_i|\big(\P B\P({A-\l})^{-1}
\big)^n|\phi\rangle.\label{2.20dg}
\ee 
Now, $\|\P B\P\|\le \e|\L|$, and $\|\P(A-\l)^{-1}\|\le(3\D|\L|/2-\l)^{-1}$.
Moreover, using the fact that $-T\le U$, we have 
$H_\L=-U+zM-\e T\le -U(1-\e)+zM$. Therefore, if $z+1/2>\e/2$, we get 
$E_0\le -U_0(1-\e)$, so that, 
$$\l=E_i+U_0<E_0+U_0+\D|\L|\le \e U_0+\D|\L|=|\L|(\e/2+\D).$$ 
Consequently, 
$\|\P(A-\l)^{-1}\|\le 2|\L|^{-1}(\D-\e)^{-1}$. Using these estimates and
\eqref{2.20} we obtain, choosing $n=\D|\L|/3$,
\be |\media{\psi_i|\phi}|\le \Big(\frac{2\e}{\D-\e}\Big)^{\D|\L|/3},\ee
which implies the desired exponential bound on $R_-$. 

By inserting the bounds we obtained on $R_+$, $R_-$ into \eqref{2.15}, we
get 
\be \media{P^{32}}^{2/|\L|}\le
e^{-\b(2\D-\e)}+\big(\frac{2\e}{\D-\e}\big)^{4\D/3}.\ee
with $\D=1/16$. By proceeding analogously, we find that $\media{P^{34}}$
admits exactly the same bound. 

The estimate for $\media{P^0}$ is much simpler: in fact the state $\phi_0$
corresponding to the projector $P^0$ is an eigenstate of $H_\L$ with energy 
$-U_0+(z+1/2)|\L|$, which together with \eqref{zeta} immediately implies
\eqref{p0}. 

Finally, we have to estimate 
$\media{P^{30}}$ and $\media{P^{20}}$. 
By using RP again we can bound both of them by $\media{P^0}^{1/2}$. This is
accomplished by repeated reflections about 
the vertical planes separating regions occupied by dimers from regions
occupied by monomers, see Fig.1. In this way, we obtain the inequalities
$\media{P^{30}}\le \media{P^3}^{1/2}\media{P^0}^{1/2}$ and 
$\media{P^{20}}\le \media{P^2}^{1/2}\media{P^0}^{1/2}$, but
$\media{P^3}$ and $\media{P^2}$ are smaller than 1, obviously. Putting
things together we obtain the desired estimate \eqref{d}, which concludes
the proof of our main result.

\section{Possible extensions and conclusion}\label{sec3}

Our model, consisting of dimers with (attractive) parallel dimer-dimer
interaction and with a `flipping' type `kinetic energy', has been shown to
have LRO at
low temperature and not too large  flipping rate. We can also allow
monomers, as long as their chemical potential is not too negative. 

Unlike other proofs of these properties, ours uses reflection positivity
(RP), 
which has the advantage of simplification and the possibility of 
easily obtaining estimates of the allowed constants. It also has
the advantage that the existence of RP allows one to make statements about 
the signs of some correlation functions and inequalities among them. 
We do not go into details about these matters in this paper, however. In the interest of simplicity we concentrated only on the existence of
LRO.

Our work can easily be extended in several ways. One is to other dimensions.
The dimers can be replaced by plaquettes in 3 dimensions, although they can
also remain as dimers.

Another extension is to give a certain amount of dynamics to the monomers
by adding a term to the Hamiltonian that annihilates a pair of adjacent
monomers and replaces them with a dimer (and the reverse, of course). 
Other kinetic terms allowing, e.g., sliding of dimers over
unoccupied sites, are not generally permitted because they
break reflection positivity. 

Extensions to other lattices are possible, too: for instance, the quantum
dimer model  on the hexagonal lattice is reflection positive. However,
the proof 
of the chessboard estimate in that case is not as immediate as on the
square lattice; see \cite{FrL} for a use of repeated reflections on the
hexagonal lattice in the context of the Peierls instability in graphene. 

\medskip

{\bf Acknowledgements.} The research leading to these results has received funding from the European Research
Council under the European UnionÕs Seventh Framework Programme ERC Starting Grant CoMBoS, grant
agreement n$^o$ 239694 (A.G.) and from  U.S. National Science Foundation
grant PHY-1265118 (E.H.L.).

\end{document}